\documentclass[aps,pre,reprint,groupedaddress]{revtex4-1}
\usepackage{graphicx}
\usepackage{dcolumn}
\usepackage{adjustbox}
\usepackage{hyperref}
\begin{document}
\newcommand{\eps}{\varepsilon}
\title{The "sugar" coarse-grained DNA model}
\author{N.A.~Kovaleva}
\email{natkov@polymer.chph.ras.ru}
\author{I.P.~Koroleva~(Kikot)}
\author{M.A.~Mazo}
\author{E.A.~Zubova}
\email{zubova@chph.ras.ru}
\affiliation{%
N.N. Semenov Institute of Chemical Physics, Russian Academy of Sciences,
4 Kosygin Street, Moscow 119991, Russia
}
\date{\today}
\begin{abstract}
More than twenty coarse-grained (CG) DNA models have been developed for simulating the behavior of this molecule under
various conditions, including those required for nanotechnology. However, none of these models reproduces the DNA
polymorphism associated with conformational changes in the ribose rings of the DNA backbone. These changes make an
essential contribution to the DNA local deformability and provide the possibility of the transition of the DNA double
helix from the B-form to the A-form during interactions with biological molecules. We propose a CG representation of
the ribose conformational flexibility. We substantiate the choice of the CG sites (6 per nucleotide) needed for
the "sugar" GC DNA model, and obtain the potentials of the CG interactions between the sites by the "bottom-up"
approach using the all-atom AMBER force field. We show that the representation of the ribose flexibility requires
one non-harmonic and one three-particle potential, the forms of both the potentials being different from the ones generally
used. The model also includes (i) explicit representation of ions (in an implicit solvent) and (ii) sequence dependence.
With these features, the sugar CG DNA model reproduces (with the same parameters) both the B- and A- stable forms under corresponding conditions and demonstrates both the A to B and the B to A phase transitions.
\end{abstract}
\maketitle
\section{Introduction}

The local DNA deformability is crucial for the DNA biological functioning: winding around histones (unwrapping and rewrapping), interactions with proteins, transcription and replication.
The deformation of the double helix is mostly achieved through three types of mobility in the DNA backbone:
concerted change of torsion angles $\alpha$ and $\gamma$ \cite{2002-alpha-gamma-Lavery}; $\zeta$ and $\varepsilon$ \cite{1993-BI-BII-Lavery} (for notations, see fig.~\ref{DNA-strand}); and ribose flexibility (the change of conformation of the sugar rings). The last mentioned type makes an essential contribution into providing the observed extreme deformability.

The local changes of sugar conformations are very common in physiological saline. In vivo, they take place in the binding of some proteins (such as TBP, SRY, LEF-1, PurR) to the minor groove of the (common) B form of double stranded DNA (dsDNA).
During this process, the minor groove widens through transitions of several sugar rings into conformations characteristic for the A-form of dsDNA  \cite{1999-MinorGrooveBindingProteins,2000-DNADeformationsExp}. Many local B to A conversions have been observed in protein and drug-bound DNA crystal complexes \cite{2000-Olson-A-DNA-with-proteins}. Particularly, such conversion takes place when an enzyme interacts with the atoms ordinarily buried within the backbone (O3', for example).
Generally, a dsDNA can assume an A-form structure when the properties of the solution and/or the amount and/or type of ions near its surface change, as well as during interaction with the partial charges on the surfaces of biomolecules. The geometric shape of the A-DNA differs from that of the B-DNA, and the mechanical properties of these helices are also different.
One can obtain DNA crystals from a solution in both the A- and B-forms depending on salt concentration, relative humidity and base
 pairs sequence\cite{2006-Historical-DNA-polymorphism}. In a solution, one can induce a B to A transition by increasing salt
concentration and/or adding ethanol to the solvent \cite{Ivanov1973,Nishimura1986}. The ethanol concentration, at which the B to A
transition occurs, depends on the nucleotide sequence \cite{1992-Ivanov-A-in-ethanol}: more C:G pairs shift the transition to
smaller ethanol concentrations. It should be noted that other characteristics of a DNA nucleotide (both structural and dynamical
ones) also depend on its base and a few neighboring bases.

Therefore, the geometry (and, as a consequence, mechanical and dynamical properties) of both the A-DNA and B-DNA forms
is a result of a complex balance of interdependent factors: torsion angles in the backbone; ribose conformations;
sequence dependent base pair stacking and pairing; electrostatic interactions of DNA with solvent molecules and with salt ions.
From the physical point of view, the understanding of this balance is equivalent to the construction of a coarse-grained
(CG) DNA model able to reproduce the needed features of the DNA behavior.

The last fifteen years witnessed the extensive development of CG models for different substances \cite{2012-Coarse-grained-for-macrochemistry}, particularly for large biomolecules \cite{2012-Coarse-grained-biomolecules,2013-Noid-bio-CG-review}. A series of regular methods for obtaining CG force fields have been developed \cite{2013-Systematic-CG-Vegt,2013-Systematic-CG-Noid}. However, the regular methods imply that the potentials are pairwise and/or have a given simple form. As we will show later, if the objective is to model the effect of the ribose flexibility on the DNA geometry, one has to use one non-harmonic and one three-particle potential, both of the form different from the potentials generally used.

Because of the complexity of the behavior of the DNA molecule and because of the different needed level of detail in various
physical situations, more than 20 CG DNA models have already been developed. There are several good reviews of these
models
\cite{2012-ModelingDNA-Levitt,2013-Papoian-review-CG-DNA,2016-MultiscaleDNA-Orozco,2016-Lyubartsev-review-on-DNA-and-histone},
we will only describe one trend in their evolution.

To reproduce the properties common to dsDNA and other semiflexible polymers, one can use the simplest worm-like chain
model with a bead consisting of two complementary nucleotides \cite{2005-worm-like-DNA}. Similar models are being exploited
in extensive simulations of long DNA molecules, for example, in the analysis of the behavior of a dsDNA confined in a
nanochannel \cite{2009-DNA-in-nanochanel1,2011-DNA-in-nanochanel2,2014-DNA-in-nanochanel3} or in simulation
of single stranded DNA molecules
(ssDNA) in DNA-functionalized spherical colloids \cite{2013-worm-in-colloid}. In this model, one usually
obtains the model parameters from comparison with the experimental data ("top-down" approach). A modification of this
model \cite{2009-Mazur-CG-bending-torsion} allows one to obtain both bending and torsional persistent lengths (consistent
with experiment). However, the modified model gives one order of magnitude lower value for the relaxation rate of bending
fluctuations \cite{2009-Mazur-CG-bending-torsion}. The reason is that the DNA bending is more sophisticated than
in the worm-like model because  on the scale less than 100nm the dsDNA behaves rather like a stack of interacting plain
nucleobases \cite{2007-CG-DNA-rigid-bases}.  The models including plain bases can represent the local inner dynamics
of the molecule more adequately, and these models are able  to take into account the sequence dependence
 \cite{2003-stack-of-bases-Olson,2003-stack-of-bases-Everaers,2007-CG-DNA-rigid-bases}.
 To derive the potentials of interactions
 between the bases, one can use the results of classical all-atom simulations or quantum chemical calculations ("bottom-up"
 approach) \cite{2010-DNA-CG-2grains-plus-ellipsoid,2010-CG-bases-Sheraga,2012-ab-initio-between-bases}.

 To simulate the dsDNA denaturation and hybridization (as well as transcription or replication), one needs at least two
 interacting chains of beads. The simplest approach is to use one interaction site per nucleotide.
 Depending on the intended application, interchain (interstrand) interactions can have different levels of complexity:
 a bead on one chain may interact with one \cite{2008-DNA-denaturation-2-chains-Sung}, three
 \cite{2014-DNA-denaturation-2-chains-Sung2,2010-two-chains-for-SNP-Kawano}
 or many beads \cite{2009-CG-DNA-renorm-Papoian,2014-Laaksonen-CGDNA} on the second chain. The models of this type
 are able to reproduce the geometrical shape of B-DNA, and, together with explicitly modelled ions, the dependence
 of the persistence length on the ionic concentration \cite{2016-Lyubartsev-review-on-DNA-and-histone}.
 These models allow to develop the regular methods for obtaining the interaction potentials
 between the beads: molecular renormalization group coarse-graining \cite{2009-CG-DNA-renorm-Papoian}, newton inversion method
 \cite{2014-Laaksonen-CGDNA}. However, to simultaneously simulate, on the one hand, the correct geometry and mechanics,
 and, on the other hand, the adequate possibility for splitting into two strands, one needs, at least, one interaction site
 for the nucleobase and, at least, one interaction site for the backbone part of the
 nucleotide \cite{2010-CG-DNA-Doye-1,2014-CG-DNA-Aksimentiev}.
 With the proper interactions between the bases, this model is elaborate enough for simulation of DNA melting, hairpin
 formation and duplex hybridization \cite{2011-CG-DNA-Doye-2}, as well as the sequence dependence \cite{2012-CG-DNA-Doye-3}.
However, the dynamics of the molecule backbone is better modelled if one uses two sites on the backbone per nucleotide:
one for the phosphate group and one for the sugar. The number of sites on the base also may be chosen to be more than one
to better describe stacking and pairing. Actually, most of the CG DNA models are of this
type \cite{2001-CG-DNA-Schatz,2005-CG-DNA-Voth,2007-CG-DNA-3-grains,2010-CG-DNA-Plotkin,2011-CG-DNA-Dorfman,2014-Scheraga-CG-dsDNA-with-bases,2015-CG-DNA-Nguyen}. For a model of this type, a force field for DNA-protein interactions
has been offered \cite{2015-CG-DNA-proteins-Sheraga}.

To deal with the same problem of the protein-DNA docking, another model of
CG DNA has been elaborated \cite{2008-CG-DNA-for-CG-proteins-Prevost} for the CG proteins earlier
offered \cite{2003-CG-proteins-Zacharias}.
In this model, the third interaction site on the backbone part of the nucleotide was
introduced (the ribose ring is divided into two beads). This CG model well reproduces the geometry of some protein-DNA complexes,
and so it seems that the shape of the CG strands is fine enough to model the needed large local deformations of the dsDNA
in the complexes. The model with the same number of beads on the backbone \cite{2010-CG-breathing-sequence-dep-Uruguay} can
well reproduce the dsDNA breathing dynamics (and melting) and its dependence on sequence specificity.
However, the natural question arises: are these strands flexible enough if the potentials of interactions
between the beads do not allow to simulate the conformational changes in the ribose rings, whereas the ribose
flexibility significantly contributes to the flexibility of the strands?

We think that the answer is 'no'. To adequately reproduce DNA local deformability, the model needs to reproduce
the ribose flexibility. It means that both the B- and A-DNA forms and the transitions between them should be present
at the corresponding physical conditions.
In this article, we propose the first CG model meeting this requirement. More precisely, using all-atom
simulations and experimental data, we justify the choice of beads on the backbone (their minimal needed number and location)
and the potentials of interactions between them. We show that the existence of the A-DNA can be provided only
if the ions are treated
explicitly. For the derivation of the model, we exploit the interactions between bases as in the AMBER
force field, but the proposed method of introducing ribose flexibility may be used with any of the CG
base interactions obtained by "bottom-up" approach
\cite{2010-DNA-CG-2grains-plus-ellipsoid,2010-CG-bases-Sheraga,2012-ab-initio-between-bases}.

We formulated the choice of the beads and derived the potentials for simulating ribose flexibility in
2010 \cite{2010-first-full-publication}. The version of the model without implementation of the ribose flexibility,
without the sequence dependence, and with implicit solvent (generalized Born approximation with the Debye-Huckel correction
for salt effects \cite{Case1999}) has been used for estimation of the B-DNA heat conductivity \cite{Savin2011}
and for investigation of dsDNA stretching \cite{Savin2013}. However, this version is admissible only when one
may neglect the ribose flexibility (under low temperatures or when the molecule can be extended but not bent and
not compressed). At temperature 300~K, the behaviour of this CG B-DNA and the all atom AMBER model significantly
differ \cite{2011-Biofizika}. Here, we present the full model (hereinafter referred to as the "sugar" CG DNA model),
including explicit representation of ions and sequence dependence.
The parameters of the model are accurately adjusted, so that it can simulate both B- and A-DNA at proper conditions,
and demonstrates both the A to B and the B to A transitions.

\section{Model construction}
For base pairs stacking and pairing, we accept the AMBER force field Parm99SB+bsc0 \cite{2000-Parm99,2006-AMBER-SB,2007-parmbsc0}.
This choice is justifiable, considering that the empirical force fields reproduce the forces acting between bases better than
any other forces in the DNA molecule \cite{2006-QM-and-AMBER-base-stacking}.
We analyze the dsDNA dynamics in solvents at 300~K in the framework of the all-atom AMBER model (for B-DNA in water -
Parm99SB+bsc0, for A-DNA in the  mixture of ethanol and water (85:15) -
Parm99 (we used the trajectory obtained  )).
Besides, we analyze the geometric changes of the dsDNA helix at A-B transition and at base-pair opening.
On the basis of these data, we divide the DNA strands into "beads" and find the potentials for interactions
between the beads. We obtain the potentials describing the ribose flexibility from
the AMBER force field (Parm99SB+bsc0) using the "bottom-up" approach. Finally, we choose the description
of the medium: the models for the ions and the solvent.
\subsection{Beads on nucleobases, stacking and pairing}
\begin{figure*}[t]
\centering
\begin{minipage}[t]{.47\textwidth}
\centering
\includegraphics[width=.98\linewidth]{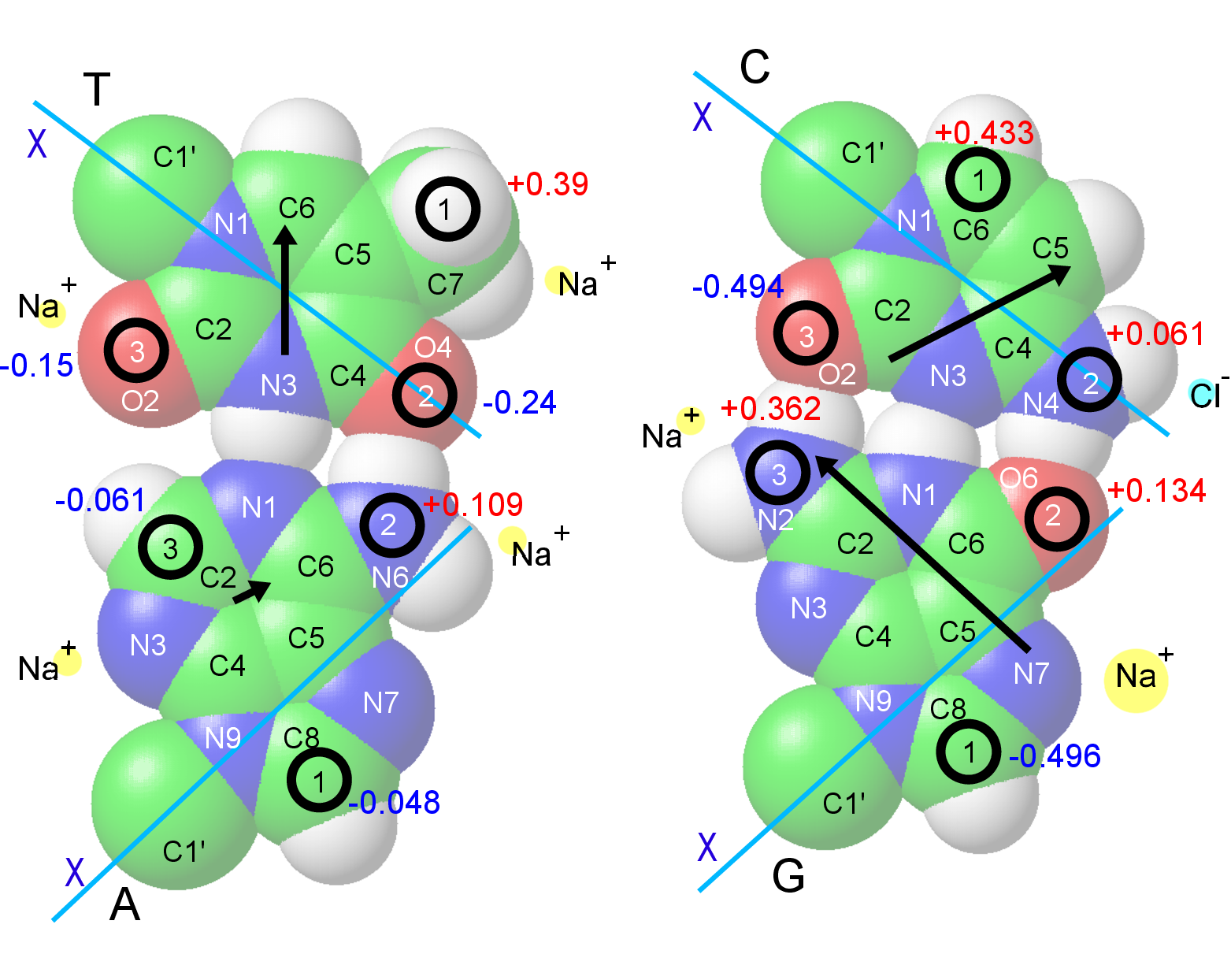}
\caption{
Coarse graining of the natural base pairs (A:T and G:C). We show the locations and and the charges of three rigidly bound beads modeling every base. The rotation axes (glycosidic bonds $\chi$ with the sugar rings) are marked.
We use common designations for DNA atoms \cite{2008-Neidle-Principles-of-DNA-structure}, and we depict their van-der-Waals radii.
The black arrows present electric dipole moments of the bases (according to the charge distribution in AMBER),
and the circles near atoms of the bases - zones around A-DNA where one can find ions Na$^+$ or Cl$^-$ with maximal probability \cite{1999-ions-around-A-DNA-AMBER-CHARMM}.
}
\label{DNA-bases}
\end{minipage} \hfill
\begin{minipage}[t]{.47\textwidth}
  \centering
  \includegraphics[width=.98\linewidth]{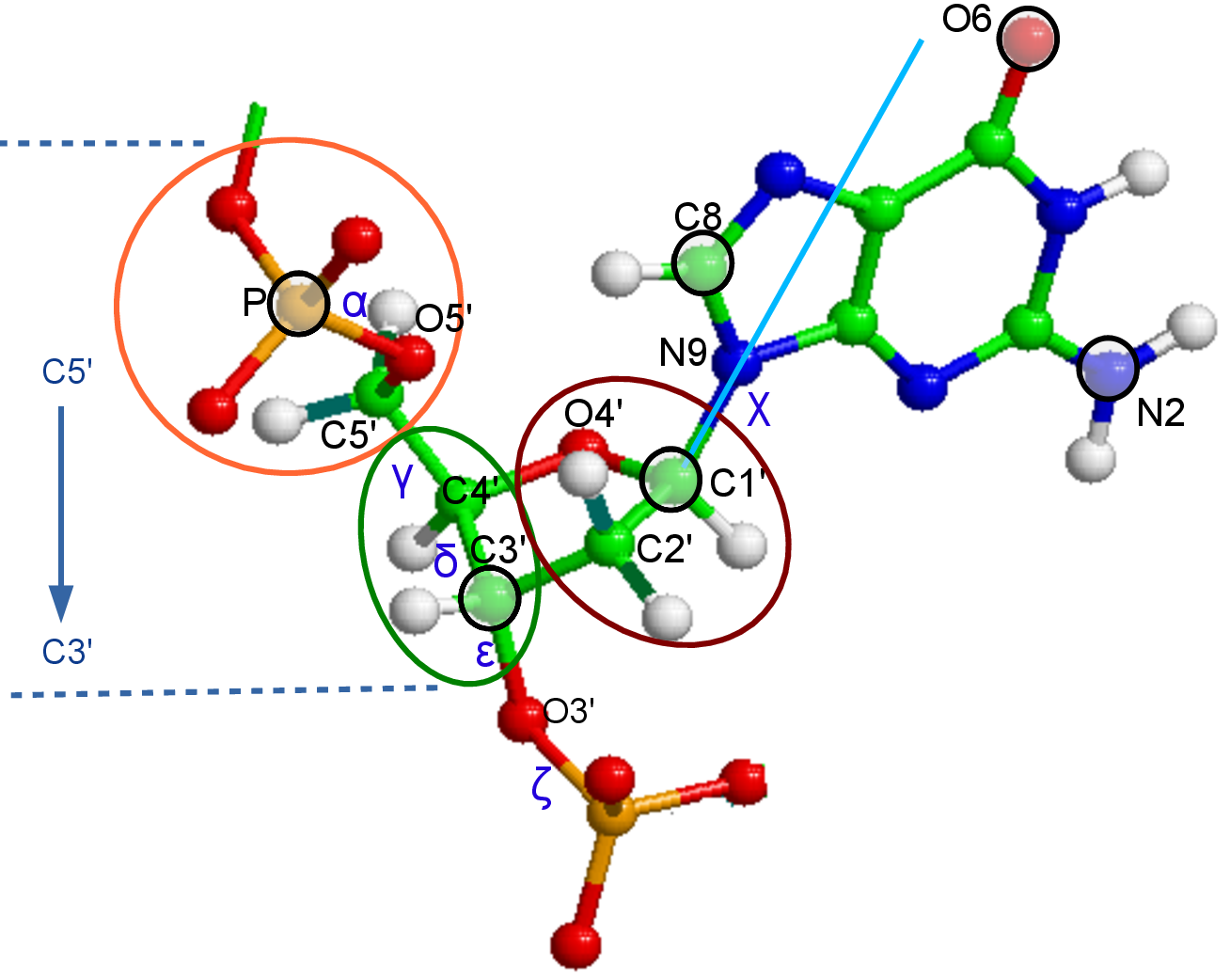}
  \caption{
  Coarse graining of the sugar-phosphate backbone. We show locations of the beads and (for the backbone)
  the groups of atoms united into the beads. We use common notations for the atoms and torsion angles
  \cite{2008-Neidle-Principles-of-DNA-structure}. A nucleobase rotates about the glycosidic bond $\chi$.
  }
  \label{DNA-strand}
\end{minipage}
\end{figure*}

We model a base as three rigidly bound beads which can rotate
around an axis coinciding with the real rotation axis of the base: the glycosidic bond $\chi$.
We placed the beads on some (heavy) atoms of the bases (see fig.~\ref{DNA-bases}).
One of the atoms was chosen on one side of the rotation axis, the two others - on the other side,
at maximum distance from the center of the base. The objective of this choice was to best approximate
the real moments of inertia of the base.
For base A (Adenine) we placed the beads on atoms C8, N6, C2; for base T (Thymine) - on atoms
C7, O4, O2; for base G (Guanine) - on atoms C8, O6, N2; and for base C (Cytosine) - on atoms C6, N4, O2 (see fig.~\ref{DNA-bases}).
Masses of the beads $m_1$, $m_2$, $m_3$ on a base X (X = A, T, G, C) were found from the two conditions: (1) equality of the
total mass of the beads to the mass of the base ($m_1+m_2+m_3=m_X$) and (2) coincidence of the mass centers of the
three beads and the base. Numerical values for masses of beads and moments of inertia of the CG and real bases are
given in table~\ref{tab1}.
\begin{table}
\caption{Masses of beads $m_1$, $m_2$, $m_3$ and moments of inertia of the real i$_{xx}$ and i$_{yy}$ and the CG I$_{xx}$ and
I$_{yy}$ bases A, T, G, C. Masses of the beads are given in a.e.m., the moments of inertia - in a.e.m.$\cdot \mbox{\AA}^2$.
}
\label{tab1}
\begin{tabular}{cccccccc}
\hline
X & $m_1$  & $m_2$ & $m_3$ & i$_{xx}$ & I$_{xx}$  & i$_{yy}$  & I$_{yy}$\\
\hline
A & $52.230$ & $28.139$ & $53.632$ & $690$   & $475$ & $1704$ & $1712$\\
T & $51.822$ & $16.204$ & $56.974$ & $584$   & $256$ & $1636$ & $1543$\\
G & $61.731$ & $34.357$ & $53.912$ & $1302$ & $800$ & $1885$ & $1858$\\
C & $39.254$ & $35.492$ & $35.254$ & $233$   & $164$ & $1344$ & $1231$\\
\hline
\end{tabular}
\end{table}

Knowing the coordinates of the three beads, one can find the coordinates of all the base atoms and compute
the energy of base pairing and base stacking as the sum of pairwise atomic electrostatic and van-der-Waals
interactions. For them, we used the AMBER force field (Parm99SB+bsc0 \cite{2000-Parm99,2006-AMBER-SB,2007-parmbsc0}).
Savin et al \cite{Savin2011,Savin2013}
used the partial interaction between the bases to decrease the computation time. In the present realization
of the model, every atom of a base interacts with every atom of the complementary base and with all
the atoms of two neighboring base pairs. It allows to simulate the A-DNA form and other deformed structures
(for example, base pair opening).

The accepted interactions between bases are computationally expensive (as compared to the other interactions
in the CG model). We used them only as a foundation to construct the adequate CG backbone and test the obtained
structure. If one uses the sugar CG model in long simulations of biological processes in the future, the described scheme
should be replaced with a CG one (derived by "bottom-up" approach \cite{2010-DNA-CG-2grains-plus-ellipsoid,2010-CG-bases-Sheraga,2012-ab-initio-between-bases}).
\subsection{Beads on backbone}
\label{choice-of-grains}
We choose the locations of all the beads on (heavy) atoms. The main principle was:
one may distort the mass distribution of the molecule, but one preserves its key geometric nodes.
We will show that, to achieve this, one needs three beads per nucleotide for the backbone (see fig.~\ref{DNA-strand} ).

The first bead was chosen on atom C1' because it is the point of suspension of the base,
and the rotation axis of the base (glycosidic bond) passes through this atom. Crystallographic data
\cite{1996-rotation-around-glicosidic-bond} show that bases rotate around this axis considerably,
while valence angles O4'C1'N1(N9) and C2'C1'N1(N9) practically do not change and fix the direction
of the glycosidic bond relative to the ribose ring.
This is the key geometric peculiarity of the connection between a base and the backbone, and we keep
this peculiarity in our model.

The second bead was chosen on phosphorus atom, and united phosphate group and three atoms (C5', H5'1, H5'2) which normally move
together with the phosphate \cite{1999-PO4CH2-together}. As we want to minimize the number of beads per nucleotide, we would tend
to restrict the backbone to the chain (...-P-C1'-P-C1'-...). However, all atom MD simulations show that the presence
of the very flexible sugar rings between the atoms P and C1' allows the dihedral angles in this chain to vary over
a very wide range,
while the dihedral angles in the chain (...-P-P-P-...) keep their mean values quite satisfactorily. To preserve this feature,
we need at least one more bead per nucleotide. Besides, we would like to build a backbone which can serve as a support
for the moving bases at their opening and in A-B transition
(which is the case for the real sugar-phosphate backbone of DNA). This purpose is also not achieved for the
chain (...-P-C1'-P-C1'-...) as the atoms C1' always move together with the bases.

\begin{figure}
\begin{center}
\includegraphics[width=0.95\linewidth] {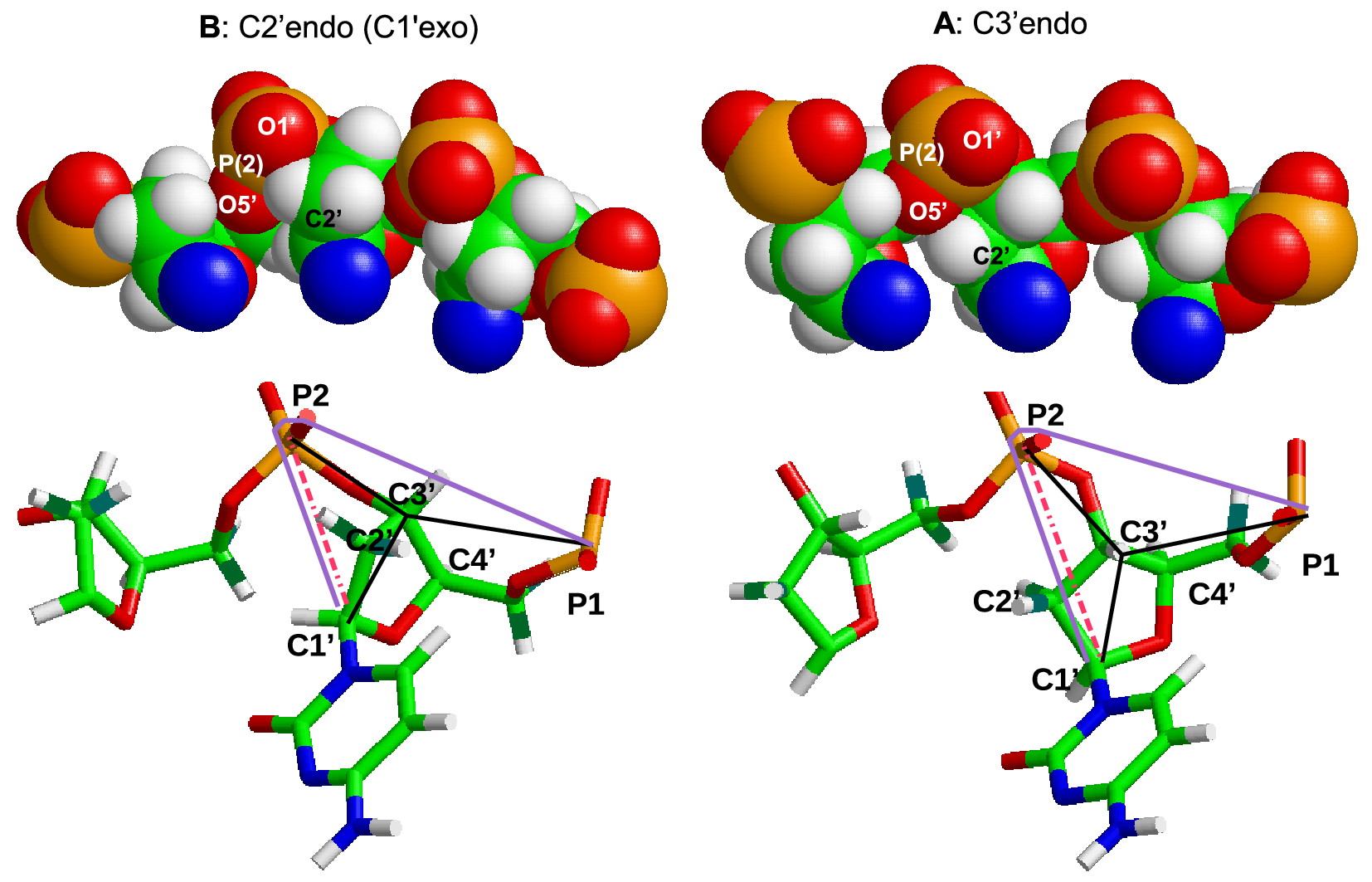}
\caption{
CG model of the ribose flexibility. The valence bonds and angles, on the one hand, and the van-der-Waals interactions,
 on the other hand, provide two stable positions of the ring relative to the phosphates exactly when the ring
 has C2'-{\it endo} (B-DNA) and C3'-{\it endo} (A-DNA) conformations: between two phosphates and under the
 chain of phosphates. Intermediate positions suffer strong steric hindrance, which results in a barrier between
 these two energy minima. When the ring changes its conformation, it has to change its position relative to the
 chain of phosphates (distance C1'P(2)). This, in turn, has to change the distance between the phosphates.
 So we have correlations: (C2'-{\it endo} $\rightleftharpoons$ small $\mid$C1'P(2)$\mid$ $\rightleftharpoons$
 large $\mid$P(1)P(2)$\mid$ $\rightleftharpoons$ B-DNA) and (C3'-{\it endo} $\rightleftharpoons$
 large $\mid$C1'P(2)$\mid$ $\rightleftharpoons$ small $\mid$P(1)P(2)$\mid$ $\rightleftharpoons$ A-DNA).
 Therefore, we choose sufficiently rigid harmonic potentials for the CG bonds C3'-P(2), P(1)-C3' and C3'-C1'.
 For the CG bond C1'-P(2), we introduce a double-well potential. The two wells correspond to two main
 conformations of the sugar ring. The correlation of the distance $\mid$P(1)P(2)$\mid$ with $\mid$C1'P(2)$\mid$
 is described by potential (\ref{equation-3-point-potential}). This potential can be symbolically depicted as
 one spring thrown from bead P(1) to bead C1' over bead P(2). For the long CG bond P(1)-C1', we use a soft spring.
}
\label{B-A-rings}
\end{center}
\end{figure}

When a ribose ring changes its conformation, the torsion angles around the bonds adjoining the ring also change,
and so does the position of the ring
relative to the two neighboring phosphate groups, as well as the distance between the groups.
The adjacent base pairs move relative to each other, and the local geometry changes from B-like to A-like (see fig.~\ref{B-A-rings}).
One can see that the atom C4' moves with the sugar ring when its conformation changes. On the contrary, the displacement
of the atom C3' is mostly caused by the necessity to change the geometric form of the chain of phosphates.
Therefore, we choose the third bead on the atom C3'.

As a result, to hold the needed form of the DNA strand, we use CG bonds, CG angles, and CG dihedral angles in the chain of beads
(...-P-C3'-P-C3'-...). To model the ribose flexibility, we add beads C1' (connected to bases)
attached to the C3' beads of this chain. The position of the CG bond C3'-C1' relative to the chain (...-P-C3'-P-C3'-...)
should have two locations divided by a barrier - which reflects two main states of the ribose.
\subsection{Ribose flexibility}
\label{choice-of-pyramid}
Ribose flexibility is modeled via double-well potential for CG bond C1'-P(2) - see fig.~\ref{B-A-rings}.
The all atom simulations also show that it is this bond of all the distances in the CG pyramid \{P(1)P(2)C3'C1'\}
that directly correlates with the ribose conformation (see fig.
(S1) in Supplemental Material \cite{SuppMat}). The correlation between the distances $\mid$P(1)P(2)$\mid$
and $\mid$C1'P(2)$\mid$ is introduced
via three-particle potential
\begin{equation} \label{equation-3-point-potential}
U=\frac{1}{2} k_P (\left|P(1)P(2)\right|+t_P\left|C1'P(2)\right|-l_{P0})^2,
\end{equation}
where $t_P>0.$
From direct geometric considerations, one could also expect a (positive) correlation between distances
$\mid$P(1)C1'$\mid$ and $\mid$C1'P(2)$\mid$.
However, as we will see from the all-atom simulations, one can use a soft harmonic potential for this CG bond:
the distance $\mid$P(1)C1'$\mid$ does not directly correlate with the sugar conformation. Indeed, there are
five valence bonds and three torsion angles between atoms P(1) and C1', and these degrees of freedom are
only weakly connected with the sugar conformation.
\subsection{Obtaining parameters of CG potentials from all-atom modeling}
\label{section-AMBER}
\begin{figure}
\includegraphics[width=0.75\linewidth] {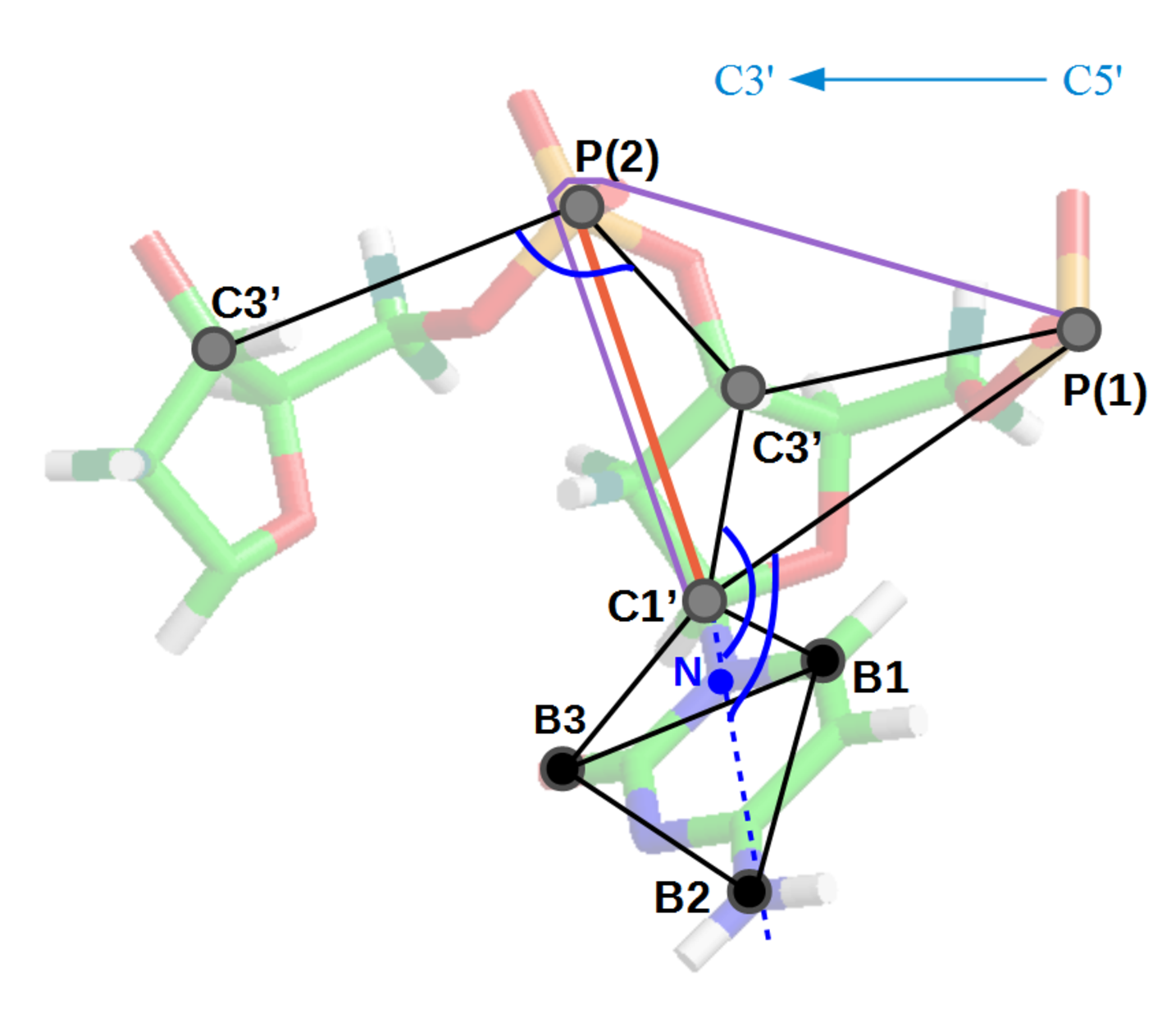}
\caption {CG bonds and angles of the sugar CG model. The double-well bond C1'-P(2) models ribose flexibility,
the length $\mid$P(1)P(2)$\mid$ correlates with the length $\mid$C1'P(2)$\mid$ (potential (\ref{equation-3-point-potential})
is symbolically depicted as the polyline P(1)-P(2)-C1').
Atom N (N1 or N9) is not one of the beads of the CG model, it is shown merely to determine the direction
of the glycosidic bond, around which the base rotates.
\label{valence-bonds-angles}
}
\end{figure}

To derive the potentials for the CG bonds and angles (fig.\ref{valence-bonds-angles}), we used two methods.
We estimated the rigidities of harmonic bonded potentials (CG bonds, CG angles, CG dihedral angles)
by the simplest Boltzmann inversion method \cite{1998-Boltzmann-inversion} for all-atom MD
trajectories of a B-form of dsDNA in water (AMBER, Parm99SB+bsc0 \cite{2000-Parm99,2006-AMBER-SB,2007-parmbsc0}) and an
A-form of dsDNA in the  mixture of ethanol and water (85:15) (AMBER, Parm99, we used the trajectory obtained
by A. Noy et al \cite{2007-A-B-Noy-Orozco}).
In this approach, the obtained rigidities take into account not only the (valence, torsion
and van-der-Waals) interactions between the DNA atoms, together with the DNA solvation. In addition, these rigidities
partly "include" several interactions which we plan
to introduce separately: base stacking; electrostatic interactions between charges on DNA, and between charges on DNA and ions.
Therefore, to verify the obtained rigidities and, in some cases, to evaluate the CG potentials which can not be derived
in such a way  (for example, in the case of the double-well potential for the CG bond C1'-P(2)), we used an another method
(method of "relaxation").

Namely, to obtain the energy of interaction between two beads at a given distance, we took an all-atom fragment
of one DNA strand
(without charges, in vacuum) between the atoms corresponding to these beads, and located these atoms at the needed distance
one from another. Then we minimized the energy of the system (in the framework of the AMBER force field Parm99SB+bsc0)
as a function of coordinates of all the rest atoms (and so we carried out the "relaxation" of the system).
The obtained value of energy was regarded as the energy
of interaction between the beads at this distance. Changing the distance between the beads, we obtained the dependence
of the energy on the distance. In this way one can evaluate potentials of "valence" bonds in a CG model.

The details of the derivation of the potentials are collected in Appendix A in Supplemental Material \cite{SuppMat}.
The description of the resulting force field is given in table \ref{constants-and-formulas}.
\subsection{Modeling DNA environment}
It is clearly seen from the all atom modeling \cite{1998-A-to-B-Beveridge-2,1998-A-to-B-Beveridge} that
most counter-ions are situated near the surface of the A-DNA, and almost in one thin layer. Closer examination
shows that the counter-ions are located mostly in the major groove of the A-DNA both in ethanol-water
mixture \cite{1997-Cheatham-MD-AB} and in a small water drop \cite{2003-Mazur-A-B-in-drop}. It allows the
phosphates on the opposite sides of the major groove to approach each other, and thus forms the characteristic
cavity of the A-DNA. It is obvious that such ion distribution can not be described in the framework of any implicit approach.
And, indeed, if one uses the generalized Born approximation with Debye-Huckel correction for salt effects, only some shift
of the DNA form from B to A can be observed when 1M of salt is added \cite{2006-Chocholousova-A-DNA-Implicit-Solvent}.
Contrary to this, with explicit media modeling in the framework of the CHARMM force field, the transition from B- to A-DNA
took place in 1.5 nanoseconds with only 0.45M of salt added \cite{1996-B-to-A-Pettitt-Sharmm}.
Therefore we introduce ions explicitly.

Explicit modeling of solvent takes lion's share of computational resources, and we know of no evidence that interaction of DNA
with solvent molecules should be treated explicitly. Therefore, all the known effects of the medium (electrostatics
and solvation), which may affect the balance of interactions in the system, are represented implicitly, via effective
potentials of interaction between the beads of the CG model, between the ions, and between the ions and the beads
of the CG model.
\subsubsection{Electrostatic interaction between phosphate beads: distance dependent permittivity}
The simplest way to model electrostatic forces between phosphates is to put negative charges (-$e$) on phosphate
beads and to introduce the Coulomb potential of interaction between these charges. Because of the small distances
between phosphates and because of the appreciable changes in the distances in B$\leftrightarrow$A transition one
has to use a distance dependent permittivity $\varepsilon(r)$ in this potential. Indeed, the dielectric constant
$\varepsilon$ is close to vacuum at small distances between the charges because there are not enough solvent
molecules between the phosphates for screening their charges. Therefore, $\varepsilon$ is normally taken equal
to 2-3 at small distances. With increasing distance $r$, $\varepsilon(r)$ is expected to reach its macroscopic value.
Starting value and slope of this curve depend on size and dipole moment of solvent molecules, as well as on the
location of the charges on the DNA molecule.

\begin{figure}[t]
\begin{center}
\includegraphics[width=0.99\linewidth] {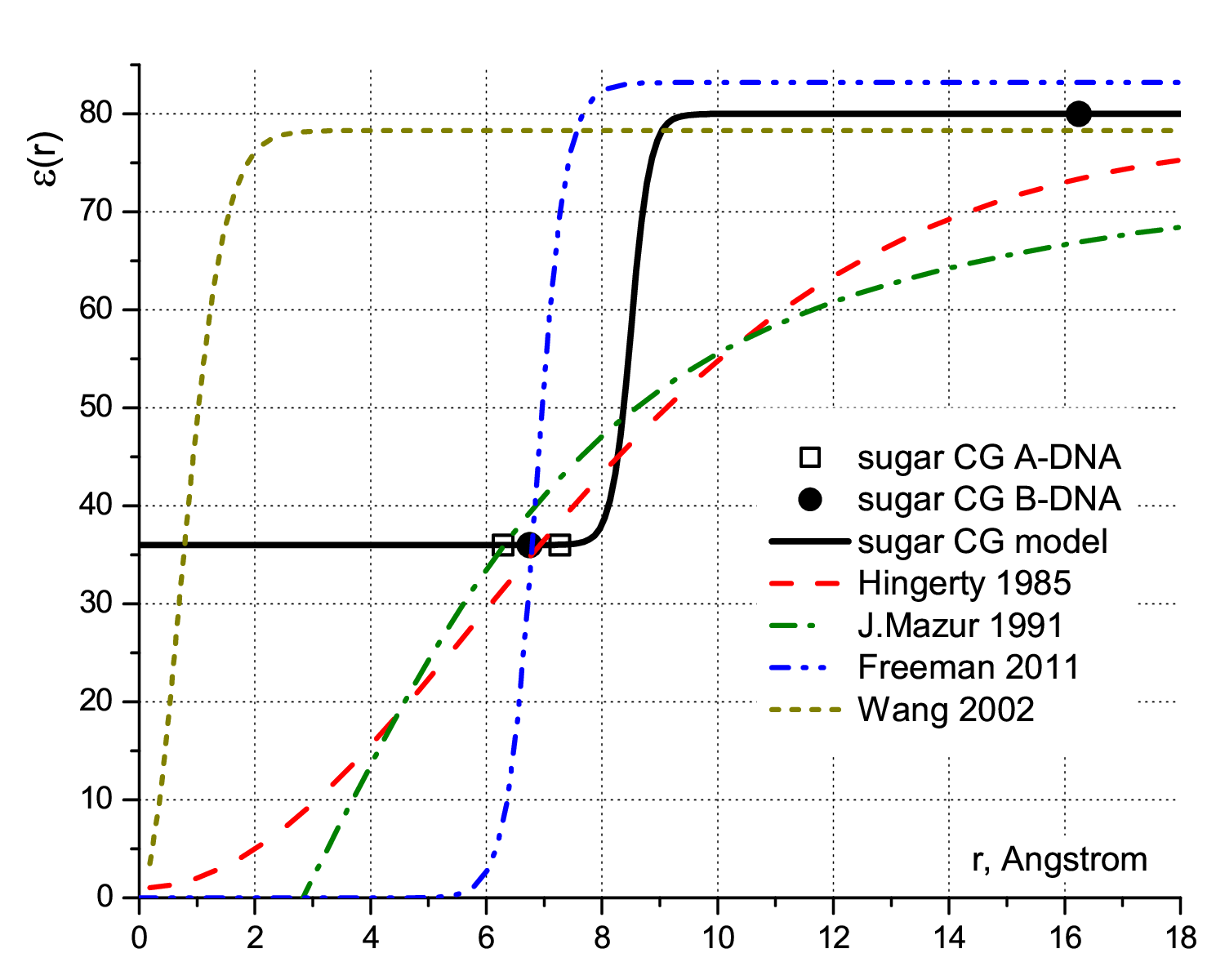}
\caption{ Screening of phosphate charges: distance dependence of permittivity $\varepsilon(r)$
in the Coulomb potential of repulsion between negative charges on phosphate beads.
We compare the dependence adopted in our model with the dependencies offered or used in other works:
by Hingerty et al \cite{1985-Hingerty-eps(r)} for charges on biopolymers, by J.Mazur
at al \cite{1991-Jernigan-eps(r)-review} and
by Wang et al \cite{2002-DNA-explicit-ions-implicit-water-Broyde} for all-atom B-DNA modeling, and
by Freeman et al \cite{2011-CG-DNA-Freeman-de-Pablo} for a CG B-DNA.
The points show the dielectric constants on MD trajectory of our sugar CG model for the nearest beads
along the strand in A-DNA and B-DNA: $\varepsilon(6.29\mbox{\AA})=36$ and $\varepsilon(6.7\mbox{\AA})=36$,
and for the nearest beads located across the major groove: $\varepsilon(7.3\mbox{\AA})=36$ (A-DNA), $\varepsilon(16.2\mbox{\AA})=80$ (B-DNA).
}
\label{n1-eps(r)}
\end{center}
\end{figure}

One may use various analytical forms for the dependence $\varepsilon(r)$
\cite{2002-DNA-explicit-ions-implicit-water-Broyde,1991-Jernigan-eps(r)-review}.
We adopted the simplest representation already used in one CG DNA model \cite{2011-CG-DNA-Freeman-de-Pablo}:
\begin{equation} \label{formula-eps(r)_new}
\varepsilon(r)=\varepsilon{_0}+\varepsilon{_1} \tanh  [\exp \left( \frac{ \alpha} {2} (r-r_{0}) \right)] ,
\end{equation}
with differing parameters: $\varepsilon_{0}=58$ and $\varepsilon_1=22$, $\alpha = 12\mbox{\AA}^{-1}$, $r_0=8.5\mbox{\AA}$.
This function is shown in fig.~\ref{n1-eps(r)}.

\subsubsection{Interactions between ions, and between ions and DNA beads: solvation effects and sequence dependence}
A-DNA can not be simulated in water with sodium counterions, even in a small box. The only exception we know
of is in a work where
a B to A transition was observed in a tiny water drop in which the surface tension contributed
to the formation of the compact
A-DNA \cite{2003-Mazur-A-B-in-drop}. Normally, one needs to add salt to a dsDNA with
counterions to observe an A-DNA \cite{1996-B-to-A-Pettitt-Sharmm}. Therefore, in the CG model,
we included explicit ions Na$^+$ and Cl$^-$ and interactions between them, and between the ions and the charges on DNA.

We also found that the charges (-e) on phosphates (e is an elementary charge) are not sufficient for the formation of A-DNA,
one needs to put partial charges on bases (on beads $B_1$, $B_2$, $B_3$ of all the bases), and so to introduce the
sequence dependence. More exactly, the A-DNA conglomerate is not stable if there are no charges on bases keeping
the sodium ions inside the major groove. Therefore, we distributed partial charges on beads (table~\ref{table-base-charge})
so that the dipole moment of every (neutral) base coincided with the moment in the AMBER force field
(Parm99SB+bsc0 \cite{2000-Parm99,2006-AMBER-SB,2007-parmbsc0}).
\begin{table}
\caption{Charges (in units of the elementary charge e) of base beads interacting with ions in solution.}
\label{table-base-charge}
\begin{tabular}{c c c c}
\hline
type of base & B$_1$      & B$_2$           & B$_3$   \\
\hline
Adenine   &     -0.048      &  0.109           & -0.061        \\
\hline
Thymine  &     0.390        & -0.240          & -0.150        \\
\hline
Guanine   &     -0.496       &  0.134          & 0.362         \\
\hline
 Cytosine  &      0.433       & 0.061           & -0.494       \\
\hline
\end{tabular}
\end{table}
One can obtain effective potential of interaction between ions in a solvent from a radial distribution function (rdf)
in an all atom simulation by several different methods
\cite{1999-Lyubartsev-DNA-ions-potentials-inverseMC,2002-Lyubartsev-DNA-ions-potentials-HNC,2006-Vegt-Na-Cl-interaction,2009-Papoian-CG-ions-renormgroup}. These methods yield the potentials of close shapes. We chose the
analytical representation of the potential function proposed by Savelyev and Papoian \cite{2009-Papoian-CG-ions-renormgroup}:
\begin{equation} \label{equation-ions-interaction}
\mathcal{V}_{ij} = \frac{A}{{r_{ij}}^{12}}
+ \sum_{k=1}^{5} D_{k} \exp^{ - C_{k} \left[r_{ij}-R_{k} \right]^2 } +
\frac{q_i q_j}{4\pi \varepsilon_0 \varepsilon r_{ij}} ,
\end{equation}
where $r_{ij}$ is the distance between $i$th and $j$th particles (between ions or between an ion and a DNA bead).
In this expression, the first term introduces excluded volume, the second term describes the shape of  peaks
and minima of the potential due to solvation, and the last term is the long-distance asymptotics: electrostatic (Coulombic) interaction between the charges.

We adopted Cl$^-$-Na$^+$ and Cl$^-$-Cl$^-$ potentials obtained by Lyubartsev
and Mar\v{c}elja  \cite{2002-Lyubartsev-DNA-ions-potentials-HNC} from all atom simulation of 0.5M NaCl electrolyte.
In the simulation the authors used the Smith-Dang's ion model \cite{1994-Smith-Dang-ions}.
Because we have few Cl$^-$ ions interacting mostly with Na$^+$ ions and one with another,
we assumed that we may use these potentials, even with a dsDNA molecule added in solution. For the interactions of Cl$^-$
ions with phosphate beads
Cl$^-$-P, we adopted the potential obtained by Lyubartsev and Laaksonen \cite{1999-Lyubartsev-DNA-ions-potentials-inverseMC},
again from all atom simulations with the Smith-Dang's ion model. We also used Na$^+$-Na$^+$, Na$^+$-Cl$^-$ and Cl$^-$-Cl$^-$
potentials derived by Lyubartsev and Mar\v{c}elja \cite{2002-Lyubartsev-DNA-ions-potentials-HNC}
as templates for potentials of interaction between ions and beads on bases. We did not put any charges
on the beads C1' and C3', and the ions interact with them only by excluded volume potential.
The details of the derivation of these potentials are in Appendix B \cite{SuppMat}.

\begin{figure*}[t]
\centering
\begin{minipage}[t]{.47\textwidth}
  \centering
  \includegraphics[width=.98\linewidth]{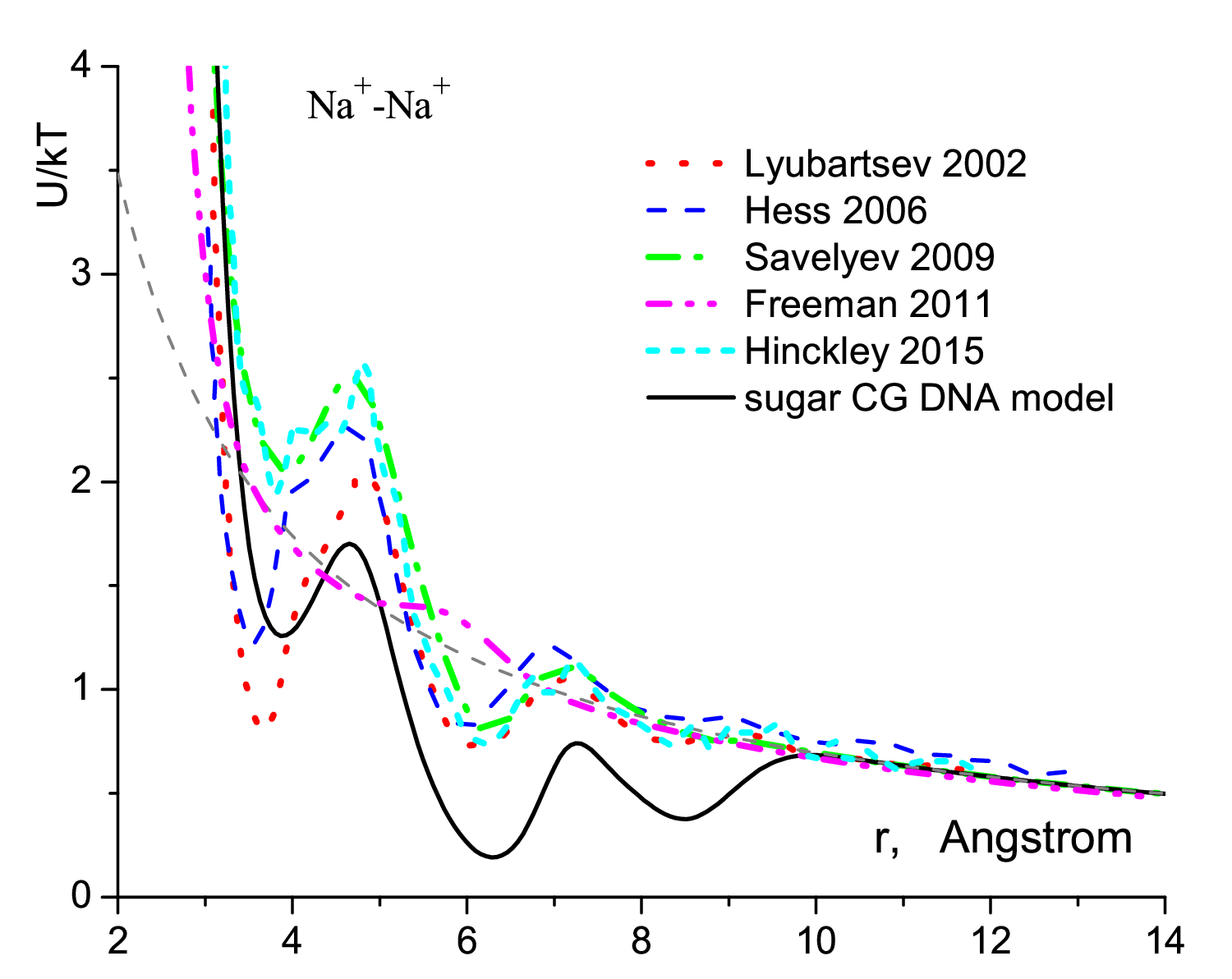}
  \caption{
  Comparison of effective potentials of sodium ions interaction. The solid line corresponds to the potential
  used in the sugar CG DNA model. We show potentials obtained by different methods from all-atom simulations
  of ions in aqueous solution in works by Lyubartsev et al \cite{2002-Lyubartsev-DNA-ions-potentials-HNC},
  Hess et al \cite{2006-Vegt-Na-Cl-interaction}, Savelyev et al \cite{2009-Papoian-CG-ions-renormgroup},
  Hinckley et al \cite{2015-CG-ions-de-Pablo}. We also show the potential
used in CG modeling of B-DNA by Freeman et al \cite{2011-CG-DNA-Freeman-de-Pablo}. Thin dashed curve corresponds
to the Coulombic interaction between the charges (the last term in equation (\ref{equation-ions-interaction})).
  }
  \label{n3-Na}
\end{minipage} \hfill
\begin{minipage}[t]{.47\textwidth}
  \centering
  \includegraphics[width=.98\linewidth]{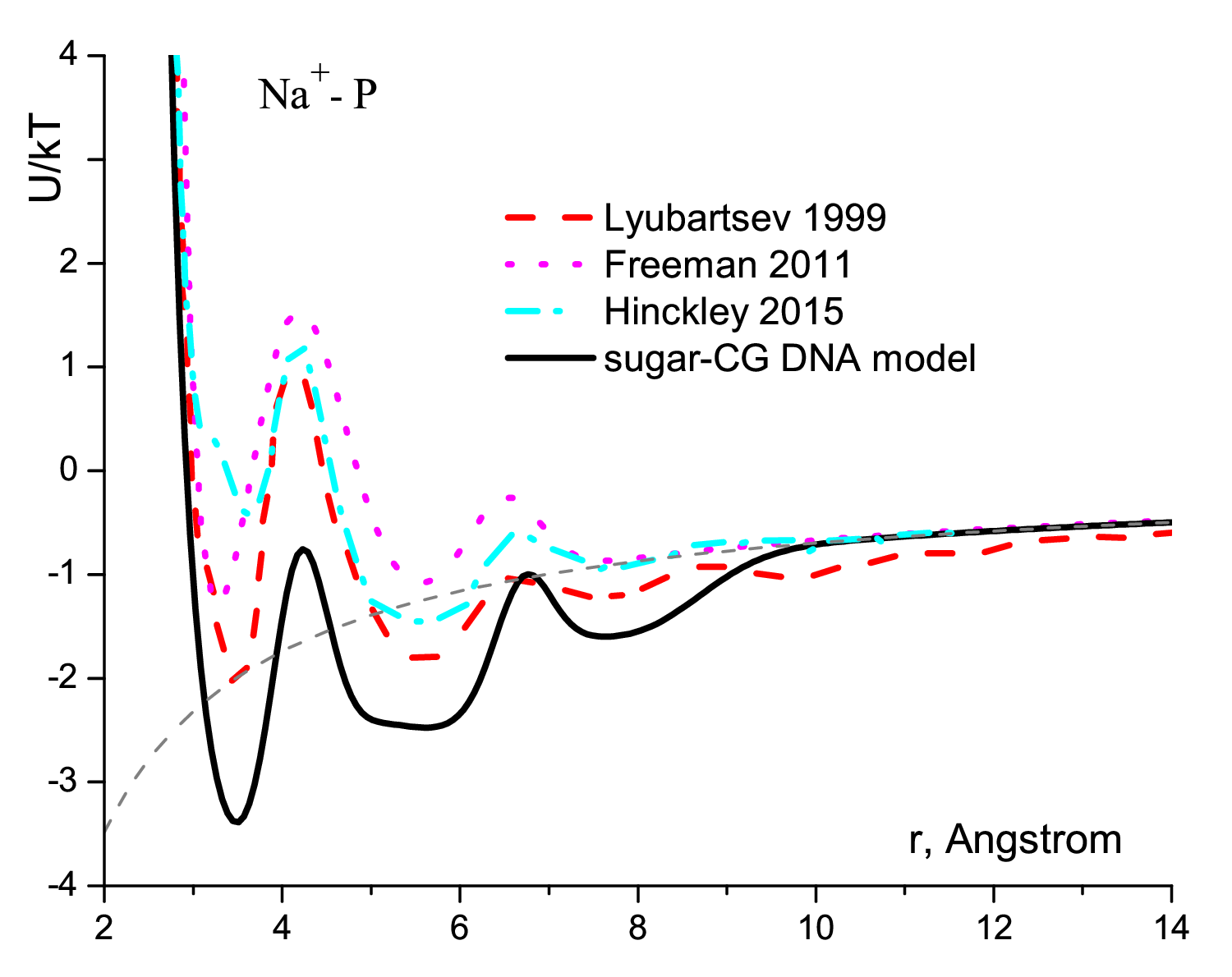}
  \caption{
Comparison of effective potentials of interaction between phosphate bead and sodium ion. The potential
used in the sugar CG model - solid line, the one obtained by Lyubartsev et al
\cite{1999-Lyubartsev-DNA-ions-potentials-inverseMC} from all-atom simulation of B-DNA - dashed line;
the one used  in CG modeling of B-DNA by Freeman et al  \cite{2011-CG-DNA-Freeman-de-Pablo} - dotted line.
Dash-dot line represents the potential obtained by Hinckley et al \cite{2015-CG-ions-de-Pablo}.
Thin dashed curve corresponds to the Coulombic interaction between the charges (the last term
in equation (\ref{equation-ions-interaction})).
  }
  \label{n3-Na-P}
\end{minipage}
\end{figure*}

Potentials of interaction between sodium ions and between sodium ions and phosphate beads were chosen so that there are
both A-DNA and B-DNA, as well as both A$\rightarrow$B and B$\rightarrow$A transitions in the model.
In figures \ref{n3-Na} and \ref{n3-Na-P} we compare these potentials with the ones obtained by different methods
or exploited in some works.
Parameters $\{A, D_{k}, C_{k}, R_{k} \}$ of the potentials used in our CG model are listed
in tables
S6-S10 in Appendix B \cite{SuppMat}. We showed some of the potential curves in
figures S3 and S4.
%
\subsubsection{Interaction of ions and beads of DNA with implicit water: coefficient of friction}
Influence of water molecules on solute molecules within implicit solvent representation is normally simulated
by Langevin equation. It provides both thermostat and viscosity. If one takes  damping (friction) coefficient
$\gamma$=70 ps$^{-1}$ for sodium ions in implicit water, their diffusion coefficient proves to be equal to the
experimental value. Prabhu et al \cite{2008-Sharp-explicit-ions-implicit-water} used the same damping coefficient
$\gamma$=70 ps$^{-1}$ for the DNA atoms fully exposed to water, while the completely buried DNA atoms had a zero
coefficient (no Langevin term). Partially exposed atoms had a damping coefficient proportional to the fraction
of its solvent exposed surface area. Chocholousova and Feig \cite{2006-Chocholousova-A-DNA-Implicit-Solvent}
accepted the value 50 ps$^{-1}$ for all DNA atoms, while Gaillard and Case \cite{2011-Case-Implicit-water-B-review} -
5 ps$^{-1}$.

The value of the damping coefficient influences the rate of relaxation processes which depend on the solvent.
This value does not seem to affect the balance
of interactions in DNA molecule. We made simulations with small friction $\gamma$=5 ps$^{-1}$ for both the DNA beads and
the ions to provide rapid system relaxation or to follow the behavior of the system for effectively longer time periods.
To observe the behavior
comparable with the all atom simulation on its timescale, we used big friction $\gamma_1=$50$ps^{-1}$ for the DNA beads and $\gamma_2=$70$ps^{-1}$ for the ions.
\section{Model description}
In the sugar CG DNA model (see fig.~\ref{valence-bonds-angles}), every one of the two DNA strands is modeled
by a zigzag of alternating beads P and C3': ...-P-C3'-P-C3'-... These beads are connected by CG bonds.
A bead C1' is linked to each C3' bead by another CG bond. This "comb" is a skeleton of the strand.
The bead C1' and the beads on the base B1, B2, B3 are connected by very rigid CG bonds C1'-B1, C1'-B3,
B1-B2, B2-B3, B2-B3.
We keep beads C1', B1, B2 and B3 in one plane by means of rigid CG dihedral angle C1'-B1-B3-B2. The three rigidly bound
beads (B1, B2, B3) almost freely rotate around glycosidic bond C1'-N(1,9) (position of the atom N(1,9) is calculated
on each step from coordinates of the beads B1, B2, B3).

To maintain the shape of the helix ...-P-C3'-P-C3'-..., we introduce, besides the CG bonds, the CG angle
C3'-P-C3' and two CG dihedral angles C3'-P(2)-C3'-P(3) and P(1)-C3'-P(2)-C3'. The position of the glycosidic bond C1'-N(1,9)
relative to the "skeleton" helix is supported by two CG
angles P(1)-C1'-N(1,9) and C3'-C1'-N(1,9). Another CG dihedral angle C1'-C3'-P(2)-C3' provides base pair opening.

Ribose flexibility is modeled by deformation of the pyramid \{P(1)P(2)C1'C3'\}. The possibility of the conformational
changes in sugar rings is provided by a double-well potential for the CG bond C1'-P(2). The length of the edge
P(1)-P(2) correlates with length of the "double-well" bond. The beads P(1) and C1' are connected by a soft CG bond.

The model system consists of a DNA double helix and explicit sodium and chlorine ions. The potential energy of the
system includes ten contributions:
\begin{eqnarray} \label{f12}
H = &E_{base} + E_{hydr-bonds} + E_{stacking} +        & \nonumber \\
      & + E_{bonds} + E_{angles} + E_{dihedrals} + & \nonumber \\
      & +E_{el} + E_{vdW} + E_{ion-DNA} + E_{ion-ion} &
\end{eqnarray}
The corresponding potential functions and the used parameters are collected in table~\ref{constants-and-formulas}.
\begin{table*}
\caption{A summary of the potential functions and parameters of the sugar CG DNA model. The order of beads
in the notation of the CG bonds and angles is their order along the chain direction (see fig.~\ref{valence-bonds-angles}).
The letter $N$ stands for atom $N1$ (or $N9$), and $C$ - for atom $C6$ (or $C8$) on bases.}
\label{constants-and-formulas}
\renewcommand{\arraystretch}{1.0}
\setlength{\tabcolsep}{2pt}
\begin{tabular}{ccccc}
\hline
\bf{Interaction}       & \bf{Potential}   &  \multicolumn{3} {c}{\bf{Constants}  }                                 \\
\hline
\hline
\multicolumn{2} {c}{CG  bonds }                    &  $r_0$ ,  \AA     &   $k_r$, kcal/(mol $\cdot $\AA$^2$)  &    \\
\hline
P-C3'                   &                                                &    4.52            &    35               & \\
C3'-C1'                &   $\frac{1}{2}k_r(r-r_0)^2$      &     2.4              &   192              &\\
C3'-P &                                                                  &    2.645          &   201              &\\
P-C1' &                                                                  &    5.4              &    28              &\\
\hline
\hline
\multicolumn{2} {c}{double-well CG bond (imitating ribose flexibility)}  &   parameter        & value   &  dimension \\
\hline
	&                                                                                             & $r_A$                        & 4.8      & \AA         \\
     &                                                                                                & $r_B$                         &4.2      & \AA          \\
     &  $U(r)=U_B(r - r_B) f(r)  +$                                                        & $r_C$                         &4.584   &\AA \\
C1'-P   &  $+ [U_A(r - r_A) + \epsilon_0] [1-f(r)] +$                            &  $\epsilon_0$              & 0         &  kcal/mol    \\
 &     $+ \epsilon_{barrier} e^{-\mu_0(r - r_C)^2}$                           & $\epsilon_{barrier} $   & -0.46    &   kcal/mol    \\
&      $U_j(r)=\frac{1}{2}K_j r^2, j=A, B$                                          &  $K_A  $                     &  63       &kcal/(mol $\cdot$ \AA$^2$)  \\
&       $ f(r)=\frac{1}{1+e^{2\mu(r-r_C)}}$                                       &  $K_B $                      &  25       &kcal/(mol $\cdot$ \AA$^2$)  \\
&                                                                                                       &  $\mu $                     &  20       & \AA$^{-1}$ \\
&                                                                                                    &  $\mu_0 $                  &  300      & \AA$^{-1}$ \\
\hline
 \hline
\multicolumn{2} {c}{CG bond correlated with ribose conformation}                                 &   parameter      & value         &   dimension  \\
\hline
                       &   $U(r_{C1'P},r_{P(1)P(2)})=$                                      &  $ k_P$       &      39        & kcal/mol   \\
        P(1)-P(2)  & $\frac{1}{2}k_P(r_{P(1)P(2)}+t_Pr_{C1'P}-l_{P0})^2$        & $ l_{P0}$       &     12.235   &  \AA   \\
         &                                                                                     &  $t_P $        &     1.27      &          \\
\hline
 \hline
\multicolumn{2} {c}{ CG angles }                               & $\theta_0$, deg & $k_{\theta}$, kcal/(mol $\cdot$ deg$^2)$        &                 \\
 \hline
C3'-P-C3'          &                                                                     &     110            &    0.017     &   \\
P-C1'-N            &   $\frac{1}{2}k_{\theta}(\theta-\theta_0)^2$    &      84            &     0.026     &       \\
C3'-C1'-N         &                                                                       &    112            &    0.032      &     \\
\hline
 \hline
\multicolumn{2} {c}{ CG dihedral  angles }                                         & $\delta_0$, deg & $\epsilon_{\delta}$, kcal/mol & comment\\
\hline
C3'-P-C3'-P            &                                                                       &     188            &   4.6         & long P-C3' bond  \\
P-C3'-P-C3'            & $\epsilon_{\delta}(1-\cos (\delta-\delta_0))$      &     194            &   4.6          &  short C3'-P bond   \\
C1'-C3'-P-C3'         &                                                                       &       13            &   3.0          & base-pair opening \\
C3'-C1'-N-C            &                                                                      &      -32            &   0.03          &  glycosidic bond\\
\hline
\hline
\multicolumn{2} {c}{ interactions in rigid bases }                      &\multicolumn{3} {c} {}                                                                   \\
\hline
         bonds                  &        $\frac{1}{2}k_r(r-r_0)^2$            &\multicolumn{3} {c} { see formula (B4)   }                                        \\
CG dihedral angle                &       $\epsilon(1+\cos \delta)$              &\multicolumn{3} {c} {and table~III by Savin et al \cite{Savin2011} }                                   \\
\hline
\hline
\multicolumn{2} {c}{hydrogen bonds and stacking interactions} &\multicolumn{3} {c} {from AMBER}\\
\hline
\hline
\multicolumn{2} {c}{ electrostatic interactions between phosphate beads}     &    parameter      & value         &   dimension         \\
\hline
              &	                                                                                                                                                           & $\varepsilon_{0}$          &   22      &    \\
$P-P$       & $\frac{q_i q_j}{4\pi \varepsilon_0 \varepsilon (r) r_{ij}} $                                                                              &$\varepsilon_1$              &   58      &      \\
            & $\varepsilon(r)=\varepsilon{_0}+\varepsilon{_1} \tanh  [\exp \left( \frac{ \alpha} {2} (r-r_{0}) \right)]  $                 &  $\alpha $                     &   0.3     & $\mbox{\AA}^{-1}$ \\
           &                                                                                                                                                                      &  $r_0$                           &   8.5     & $\mbox{\AA}$ \\
\hline
\hline
 \multicolumn{2} {c}{ van der Waals interactions between skeleton beads}                  & $\sigma_i$, \AA   & $ \epsilon_i$ , kcal/mol  &     \\
\hline
 $P$          &  $4\epsilon_{ij} \left[ \left(\frac{\sigma_{ij}}{r}\right)^{12}- \left(\frac{\sigma_{ij}}{r}\right)^6 \right] $           &  2.18              &     0.23              &                           \\
 $C3'$       &     $  \sigma_{ij}=(\sigma_i+\sigma_j)/2$,   $\epsilon_{ij}=\sqrt{\epsilon_{i}\epsilon_{j}} $                                &  2.0               &    0.115             &                           \\
\hline
 \hline
\multicolumn{2} {c}{interaction of Na$^+$ and Cl$^-$ ions with charged beads } &  \multicolumn{3} {c}{  }         \\
\multicolumn{2} {c}{of DNA and one with another  } &  \multicolumn{3} {c}{  }         \\
\hline
&  &  \multicolumn{3}  {c} { $q_{Na^+}$=+$e$, $q_{Cl^-}$=-$e$, $q_{P}$=-$e$,}         \\
&    $\frac{A}{{r_{ij}}^{12}}+ \sum_{k=1}^{5} D_{k} \exp^{ - C_{k} \left[r_{ij}-R_{k} \right]^2 } +
\frac{q_i q_j} {4\pi \varepsilon_0 \varepsilon r_{ij}} $   &  \multicolumn{3}  {c} {  charges on beads of bases see in table~\ref{table-base-charge};}         \\
&                                                                                                    &  \multicolumn{3}  {c} {  A, D$_k$, C$_k$, R$_k$, $\varepsilon$ are in tables S6-S10
}
\\
\hline
\hline
\multicolumn{2} {c}{ interaction of ions with uncharged beads }                 & $\sigma$, \AA   & $ \epsilon$ , kcal/mol  &     \\
\hline
Na$^+$ with C1'                       &  $ \epsilon \left(\sigma / r\right)^{16}$     & 3.5                    &                           0.369   &     \\
Na$^+$ with C3'                       &  $ \epsilon \left(\sigma / r\right)^{12} $    & 3.2                    &                           0.369   &     \\
Cl$^-$ with C1'                       &  $ \epsilon \left(\sigma / r\right)^{16}$     & 3.3                    &                           0.369   &     \\
Cl$^-$ with C3'                       &  $ \epsilon \left(\sigma / r\right)^{12} $    & 3.3                    &                           0.369   &     \\
\hline
\end{tabular}
\end{table*}

The term $E_{base}$ describes the energy of deformation of rigid bases.
The terms $E_{hydr-bonds}$ and $ E_{stacking}$ stand for energy of hydrogen bonds between
complementary bases and for base pairs stacking, correspondingly. We recalculate the coordinates of all nucleobase atoms
on each step and compute these terms using the all atom force field AMBER
(Parm99SB+bsc0 \cite{2000-Parm99,2006-AMBER-SB,2007-parmbsc0}).

The terms $E_{bonds}, E_{angles}, E_{dihedrals}$ describe energy of deformation of CG bonds,
CG angles and CG dihedral angles on the strands of the CG DNA.
Equilibrium values of the angles and the bonds, not pertaining to the ribose flexibility, were chosen equal
to the values in A-DNA. For the rigidities, we chose the maximal values. Two wells in
the double-well potential of the bond C1'-P(2) were made of equal depth (see fig.~S2
),
contrary to the curve obtained using the AMBER force field (Parm99SB+bsc0). We discuss the reason for these choices
in section \ref{Discussion}.

Coulombic interactions $E_{el}$ between charged phosphate beads have distance dependent permittivity
(see formula (\ref{formula-eps(r)_new})). We introduce van der Waals interactions $E_{vdW}$ for the beads $P$
and $C3'$ not interacting via bonded potentials.

Interaction of ions with DNA $E_{ion-DNA}$ includes interactions with charged phosphate beads $P$  and beads on bases
and with uncharged beads (C1', C3'). In the present realization of the model, we introduce sequence dependence: the
charges on beads of a base depend on the type of this base.
Interactions of ions one with another $E_{ion-ion}$ and with charges on DNA (phosphate beads and beads on bases) $E_{ion-DNA}$
take into account solvation effects (besides direct Coulomb force).

The influence of water on DNA and on ions is described implicitly, by Langevin equation. Damping constant (friction)
is taken to be equal to 5ps$^{-1}$ for system relaxation and
test calculations. Productive runs were made with $\gamma$=5ps$^{-1}$, as well as with $\gamma_1$=50ps$^{-1}$ for
the DNA beads and $\gamma_2$=70ps$^{-1}$ for the ions.
\section{Model testing: A-DNA, B-DNA and transitions between them}
The best test of adequacy of the representation of the ribose flexibility is the existence of both the B- and A-DNA forms
at the corresponding conditions.
We found two equilibrium states (A-DNA and B-DNA) of the system by its energy minimization (from different initial states at corresponding boundary and initial conditions).
After this, we compared CG MD trajectories of both these forms with the ones obtained in all atom simulations. Both A to B and
B to A phase transitions were observed under corresponding conditions.
\subsection{Energy minimization: A-DNA and B-DNA.}
\label{energy-minimization}
To obtain the ground states for B- and A-DNA, we started from the all atom MD configurations of a dsDNA in water and
in mixed ethanol/water (85/15) solution correspondingly. We put the beads and ions of our CG model on the places
of the corresponding atoms and ions of the all atom system.
We also added 16 Na$^+$ and 16 Cl$^-$ ions for the A-DNA. For more adequate comparison with the A-DNA, we studied
B-DNA not only in combination with
counterions (which is common), but also with the same amount of additional salt as for the A-DNA. The additional
ions were placed randomly in the unoccupied area of the
computational cell.

We modeled B-DNA in a large reservoir: a cube 60x60x60\AA.
In this volume, the 32 salt ions give the molar concentration 0.12M, very close to the one of physiological saline.
For A-DNA, we chose a small volume so that, on the one hand, the ions could not go too far from the molecule,
and, on the other hand, the energy of interaction between the chlorine ions and
the phosphate beads was not too high. The optimal reservoir proved to be a cylinder with diameter
18.5\AA\ and height 30\AA. In it, the salt concentration was 0.8M.

The energy of the CG system was minimized by the method of conjugate gradients: first the ions, and then the whole system.
As a result, we obtained both CG A- and B-DNA conformations (in corresponding computational cells) which we later used as initial configurations for MD simulations.
\subsection{MD simulations of A-DNA, B-DNA, and phase transitions between them.}
\label{section-CG-MD}
We modeled the CG dsDNA in a closed computational cell (using the reflecting boundary conditions)
with a Langevin thermostat. The initial configurations
of the system were obtained by energy minimization described in the previous section.

The initial system relaxation was being done in two stages. First, during one nanosecond, we carried out
a relaxation of the ion atmosphere with the DNA molecule kept immobile.
Then, the whole system was relaxing during the next 0.5 nanoseconds. During all the relaxation process, the friction for all the beads and ions was $\gamma=$5ps$^{-1}$. The productive runs were carried out both with the small friction $\gamma=$5ps$^{-1}$, and with the large friction $\gamma_1=$50ps$^{-1}$ for the DNA beads and $\gamma_2=$70ps$^{-1}$ for the ions.

\begin{figure}[t]
\begin{center}
\includegraphics[width=0.8\linewidth] {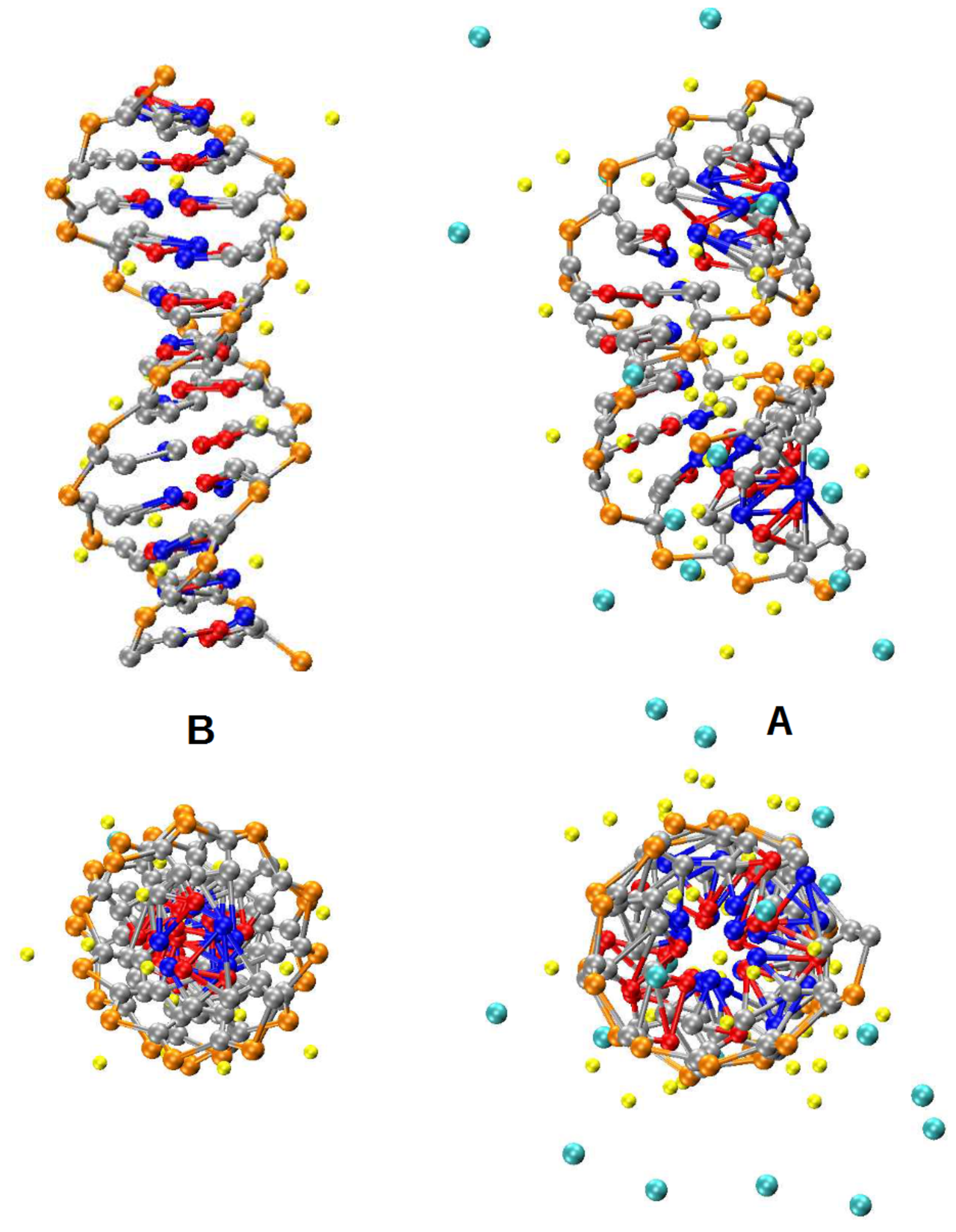}
\caption{
Frames from the trajectories \cite{A-and-B-movies} of sugar CG A-DNA (on the right) and B-DNA (on the left). Temperature is 300~K.
Sodium ions are yellow, chlorine - cyan.
}
\label{ABpictures}
\end{center}
\end{figure}

We followed the dynamics of the both forms, A- and B-DNA, at temperature 300~K up to 8 nanoseconds with small friction
and up to 18 nanoseconds with large friction, which allows to consider the forms stable at the corresponding conditions.
The B-DNA is stable in both the simulations: with counterions and in physiological saline.
The obtained stable configurations are shown in figure \ref{ABpictures}.
We compared the behavior of the A- and B-DNA in our sugar CG model and in the all atom AMBER force field
in Appendix C in Supplemental Material \cite{SuppMat}. Tables S11 and S12
list the lengths and the angles.
As one should expect, our model proved to be stiffer than the all atom one (we accepted the maximal
rigidities observed in all atom simulations), and the angles in both the A- and B-DNA are closer
to their values in the all-atom A-DNA (so we set them). The only exception is dihedral angle $\angle$C3'C1'NC,
for which the prescribed magnitude was (-32$^0$), while the observed one - (-50$^0$) in the A-DNA and (-40$^0$)
in the B-DNA. Interestingly, that the value for the A-DNA is in excellent agreement with the crystallographic
value (GLACTONE \cite{1998-GlactoneGeorgia}): see table~S4
.
As regards the ribose flexibility, the CG model imitates the all-atom A-DNA very closely, including sequence
dependence (see fig.~S5
). The sugar CG B-DNA is stiffer than the all atom one, and has lower population
in the area between C2'-endo and C3'-endo (fig.~S6
).

When we put the A-DNA into the large reservoir, we saw the A to B phase transition within 2 nanoseconds \cite{A-and-B-movies}.
The B-DNA transformed into A-DNA in the small reservoir during 6 nanoseconds (after waiting for 14 ns). 
We plan to study both the transitions more closely in the next work.

\subsection{A-DNA and ion-DNA interactions}
\label{section-ions-discussion}

A-DNA can exist only if sodium ions can assemble in the major groove so that their electrostatic interaction with the
phosphate beads (and with the nearest chlorine ions) will give the gain in free energy greater than the loss
in entropic contribution because of the ion clustering (see the balance of interactions in CG DNA forms in
fig.~S7
). For the stabilization of this positively charged cluster between two rows of
negative charges, the presence of a solvent in the major groove is crucial.

To adequately model this balance, we very precisely chose the positions and the widths of the minima of the
effective solvent-mediated potentials. We built our potentials for the Na$^+$-P and Na$^+$-Na$^+$
pairs so that the CG radial distribution functions g(r) were as close as possible to the all atom ones,
especially in what regards the positions of the minima.
The agreement between the CG rdf and the all atom one for the pair Na$^+$-P in the A-DNA is almost ideal
(see fig.~S8
).
For the pair Na$^+$-Na$^+$ it was impossible because we simulated A-DNA in water, and not in mixed ethanol/water
solution. A-DNA with counterions in water does not exist without additional salt. Therefore we had many more sodium ions
in the computational cell, and, correspondingly, in the major groove. It had to lead to the substantial rise
of the first peak as compared with the second one (see fig.~S10
),
which usually takes place with increase of the salt concentration \cite{2011-Shen-Vegt-rdf-ions}.
However, our Na$^+$-P and Na$^+$-Na$^+$ pairs are more sticky than for the default AMBER ions
(Aqvist's cations \cite{1990-Aqvist-ions}). One can
see that from rdfs for B-DNA shown in fig.~S9 and S11
.

Because of the evident problems of choice of ion parameters in the framework of additive, nonpolarizable and
pairwise potentials, there are several different sets of ion parameters in the all atom force fields.
Aqvist's cations and Dang's Cl$^-$,
which were used in AMBER by default, give the artefact of formation of stable ion pairs and even salt crystals
at moderately low concentrations (below their solubility limit). Other sets of the parameters result in rdfs
greatly differing in shape \cite{2009-Noy-Orozco-rdf-ions}. To provide the agreement of the pressure inside the DNA
arrays with experiment, one has to introduce additional corrections to the ion parameters
\cite{2012-NBFIX-for-ions-Aksimentiev}.

As compared with the default AMBER ions, our CG model gives for the pair Na$^+$-P a very high first peak (see fig.~S9
),
which means that our Na$^+$ ions are more often located near phosphates, and, consequently, one to another
(see fig.~S11
). The rdf for the Na$^+$-P pair for B-DNA in our model mostly resembles
the corresponding rdf for Cheatham's ions
\cite{2008-Cheatham-ions-parameters}. When compared to the others, Cheatham's sodium ions are much more often
near phospates,
avoiding chlorine ions and each other \cite{2009-Noy-Orozco-rdf-ions}. This feature seems to be
a drawback leading to over-neutralization of the DNA. However, it proved to be \cite{2015-Case-Cheatham-ions-around-DNA} that
the number of Cheatham's sodium ions well agrees with ion counting experiments at low salt concentrations,
and at high concentrations ($>$0.7M) is even less than in experiment. Therefore, we can regard our effective potentials for
ion interactions as trustworthy.
\section{Discussion}
\label{Discussion}\
We built our CG representation of the ribose flexibility on the basis of the all atom force field AMBER
(Parm99SB+bsc0 \cite{2000-Parm99,2006-AMBER-SB,2007-parmbsc0}). Starting from it, we faced the problem
that our CG DNA can assume a B-form structure at almost every reasonable set of parameters,
while balancing interactions in the A-DNA required some efforts.

First, we supposed that an A-DNA can exist without placing partial charges on the bases, i.e. without introduction of
sequence dependence. But that proved to be impossible. At temperature only as high as 300~K, the conglomerate of A-DNA proved
to be unstable, the charges on the borders of the major groove were insufficient to keep the ions inside this groove.

Secondly, the potentials and the constants, derived from the AMBER force field, required corrections. Namely,
for all the CG bonds and the CG angles (except for C1'P and P(1)P(2) connected with ribose flexibility) we used
equilibrium values of the all-atom A-DNA and maximal rigidities observed
in the all-atom simulations. For the double-well potential of the CG bond C1'P we lower the A-minimum
to the level of the B-minimum (see fig.~S2
\cite{SuppMat}). In this connection, one can remember that
to provide a spontaneous B to A transition of d[CCAACGTTGG]$_2$ sequence in 85\% ethanol solution in the framework
of the AMBER force field, the authors had to make "reduction of the V2 term in the O-C-C-O torsions from 1.0
to 0.30 kcal/mol to better stabilize the C3'-endo sugar pucker" \cite{1997-Cheatham-MD-AB}.

Finally, we had to exploit such effective potentials between sodium ions and phosphate beads and between
sodium ions one with another that resulted in a rdf for the pair Na$^+$-P very close to the rdf
for Cheatham's ions (see fig.~S9
), and not to the rdf for default AMBER ions (for more details,
see section \ref{section-ions-discussion}).
Evidently, the necessity of these corrections is a result of the long known "B-philia" of the AMBER force field
\cite{1997-Feig-Experiment-vs-Force-Fields}. Indeed, the B to A
transition at high salt concentration has been demonstrated \cite{1996-B-to-A-Pettitt-Sharmm} for the "A-philic"
CHARMM force field in 1996, while
for the AMBER force field this transition takes place only in a tiny drop of water \cite{2003-Mazur-A-B-in-drop}.
In it, the compact A-DNA is stabilized by surface tension. The additional salt results only in salt crystallization,
instead of the B to A transition.

After the described fitting, we have obtained both an A-DNA and a B-DNA at the corresponding conditions,
as well as both  A$\rightarrow$B and B$\rightarrow$A transitions.
\section{Conclusions}
\label{Conclusion}
We saw that the offered CG representation of the ribose flexibility in the sugar CG DNA model is adequate, as it
provides the geometric possibility of existence of both the A- and B-forms of dsDNA. The proposed scheme
is obtained by physically clear "bottom-up" approach and therefore should work equally well in both ss- and dsDNA.
This scheme seems to be the last needed component to adequately represent the deformability of a CG DNA strand,
and, consequently, the CG dsDNA molecule.

We have shown that to obtain the correct balance of
interactions in A- and in B-DNA at the corresponding conditions (high and low salt concentrations),
one should explicitly introduce charges on the nucleotides and
salt ions. Besides the charges on the phosphates, one should place the partial charges on the beads of the bases.

As is easy to see, our method of "relaxation" which we used to derive the double-well potential describing ribose
flexibility is a variant of the "derived coarse graining" \cite{2013-Systematic-CG-Vegt},
only, for the interaction between the beads, we adopted the minimum energy of all possible all-atom configurations,
instead of the mean value. This is justified because, contrary to the case of liquids, the configurations with high
energies are highly improbable on the CG time scale. The rigidities of the CG bonds and angles, although obtained
by the Boltzmann inversion method (belonging to the "parameterized coarse graining" which results are normally
state-point dependent), should not have noticeable temperature dependence, because we used relatively
"fine" mapping (each bead corresponds to a relatively small number of atoms) \cite{2015-entropy-in-CG-models}.
The electrostatic DNA-ion and ion-ion interactions contribute much more than the bonded
interactions to the entropic component
of the many-body potential of mean force. These potentials include solvation effects and are highly
temperature-dependent (we adopted the parametrization for T=300~K).

The proposed CG realization of the ribose flexibility is computationally cheap.
Together with CG interactions between bases, the sugar CG DNA model allows to promptly check physical
hypotheses in extensive simulations of long DNA molecules.
First of all, the sugar CG model can be applied for modeling of large mechanical
deformations of long DNA molecules, and not only for simulation of in vitro experiments. The charges
on the bases (depending on the base type)
allows one to use the model for studying DNA-protein interactions, including the interactions with CG proteins.
A small change of the model enables base openings, and offers the possibility to simulate DNA denaturation,
and to investigate transcription and replication. As the model includes explicit ions, one can model electrostatic
interactions of DNA and DNA-protein complexes with different types of ions in different solvents.
So, for the present, the model seems to be universal: it includes
all the needed features to be employed for any application in biophysics and nanotechnology.
\begin{acknowledgments}
We thank prof.~Alexey Onufriev who obtained an all-atom trajectory of B-DNA
in water for us; prof.~Modesto Orozco and Dr.~Agnes Noy who kindly granted us the trajectory of A-DNA
\cite{2007-A-B-Noy-Orozco}; and U. Deva Priyakumar who sent us .pdb-files of DNA molecule with an opening base
\cite{2006-MD-base-flipping}. For MD simulations, we used (properly modified) program
written by Dr.~A.V.~Savin.
The simulations were carried out in the Joint Supercomputer Center of Russian Academy of Sciences.
The work was supported by the Russian Science Foundation (award 16-13-10302).

\end{acknowledgments}
\raggedright
\bibliography{sugarDNAbib}
\end{document}